\documentclass[twocolumn, prl, aps, superscriptaddress, longbibliography,showpacs,amsmath,amssymb,floatfix]{revtex4-1}

\usepackage{changes}
\usepackage{graphicx}% Include figure files
\usepackage{dcolumn}% Align table columns on decimal point
\usepackage{bm}% bold math
\usepackage{graphicx}                                   
\usepackage{amssymb}
\newtheorem{theorem}{Theorem}
\usepackage{amsmath}
\usepackage{epsfig}
\usepackage{xcolor}
\usepackage{tabu}
\usepackage{mathtools}
\usepackage[colorlinks,linkcolor=blue,anchorcolor=blue,citecolor=blue,urlcolor=blue]{hyperref}
\usepackage{physics}
\usepackage{float}
\usepackage{diagbox}
\usepackage{inputenc}

\begin{document}

\title{Universal asymptotic solution of the Fokker-Planck equation with time-dependent periodic potentials}
% The old title
% \title{Boltzmann weight in Brownian motion with time-dependent periodic potentials}
\author{Boxuan Han}
\affiliation{Key Laboratory of Quantum Information, University of Science and Technology of China, Hefei 230026, China}
\affiliation{Anhui Province Key Laboratory of Quantum Network, University of Science and Technology of China, Hefei 230026, China}
\author{Zeyu Rao}
\affiliation{Key Laboratory of Quantum Information, University of Science and Technology of China, Hefei 230026, China}
\affiliation{Anhui Province Key Laboratory of Quantum Network, University of Science and Technology of China, Hefei 230026, China}
\author{Ming Gong}
\email{gongm@ustc.edu.cn}
\affiliation{Key Laboratory of Quantum Information, University of Science and Technology of China, Hefei 230026, China}
\affiliation{Anhui Province Key Laboratory of Quantum Network, University of Science and Technology of China, Hefei 230026, China}
\affiliation{Hefei National Laboratory, University of Science and Technology of China, Hefei 230088, China}
\affiliation{Synergetic Innovation Center of Quantum Information and Quantum Physics, University of Science and Technology of China, Hefei 230026, China}
\date{\today}

\begin{abstract}
Brownian motion, as one of the most fundamental concepts in statistical physics, has everlasting interests in interdisciplinary fields in the past century.  Although this motion with static potentials have been widely explored,  its physics in time-dependent periodic potentials are far less well understood.  Here we generalize this motion to the realm of time-dependent periodic potentials, showing that the asymptotic solution of the
probability distribution function (PDF) can  have  a universal form, that is, a Boltzmann weight multiplied by a Gaussian envelope function.   We derive a partial equation for this Boltzmann weight  and demonstrate that many different potentials can give the same Boltzmann weight. We first present an exact solvable model to illustrate the validity of our solution. For the periodic potential with a time-dependent tilt potential, we can determine the Boltzmann weight by numerical solving the partial equation. These results are confirmed by solving the Langevin equation numerically. With this idea, we can determine the asymptotic solution of the Fokker-Planck equation,  in which the entropy satisfies the thermodynamic law.  Our results can have wide applications, including quasi-periodic potentials, two-dimensional potentials and even with models with inertia, which should greatly broaden our perspective of Brownian motion. 
\end{abstract}
 
\maketitle  

{\it Introduction}: Brownian motion, which was first used to study the erratic random motion of microscopic particles (such as pollen) suspended in a fluid, is one of the most profound concepts in statistical physics \cite{einstein1905molekularkinetischen,langevin1908theorie,Chandrasekhar1943Stochastic,bian2016111years,Tongcang2010Measurement}. It is related to interdisciplinary fields, including physics, mathematics, chemistry, biology and finance {\it etc.} \cite{kramers1940brownian,hanggi1990reaction,Reimann2002brownian,hanggi2009artificial,Elgeti2015Physics,Zaburdaev2015Levy}, and even though has been proposed more than one century ago,  this model continues to be an active research frontier in these fields in theory and experiment \cite{kohler2021tracking,bousige2021bridging,Wang2025Observation,Antonov2022Solitons,Mishra2025Phase,Bebon2025Thermodynamics}.  Although the probability distribution function (PDF) in some simple potentials can be derived from the Fokker-Planck  equation \cite{Fokker1914Die,Planck1917Sitzungsberichte,Risken1996The}, its asymptotic behavior  comparatively recently been studied in more general potentials has not yet been well understood \cite{defaveri2023brownian,sivan2018probability}. 

For the above problems, we are motivated by the results in the previous literature with time-independent periodic potentials, revealing two stimulating features. Firstly, the diffusion constant $D^*$ is well-defined and can be obtained by the Lifson-Jackson (LJ) formula \cite{lifson1962self,Festa1978Diffusion,gunther1979mobility,weaver1979effective}; and secondly the PDF can be divided into the Boltzmann weight multiplied by a Gaussian envelope.  We have recently generalized this result to the quasi-periodic potentials with an external linear potential \cite{SangYang2025Brownian,gong2025proof}, yielding the similar PDF.  Further, the recent studies of Brownian motion with time-dependent field \cite{Gang1996Diffusion,Marchenko2018Enhanced,Marchenko2023Temperature,Marchenko2025Giant} also show that $D^*$ is well-defined, where there should be a Gaussian envelope in the PDF.  In this way, the determination of the Boltzmann weight should be an important task to fully determine the asymptotic behavior of the Fokker-Planck equation. 

In this manuscript we extend the previous knowledge of Brownian motion to the realm of time-dependent periodic potentials $U(x, t)$ and show that the asymptotic PDF of the Fokker-Planck equation can be written in the following universal form
\begin{equation}
p(x, t) = \exp(\varphi(x, t)) \mathcal{G}(x, t),
\label{eq-varphi}
\end{equation}
where $\varphi(x, t)$ has the same spatial period as the time-dependent potential $U(x, t)$, and $\mathcal{G}(x, t)$ is the Gaussian profile characterized by the effective diffusion constant $D^*$ and time-dependent center $x_c$.  We derive a partial differential equation for the Boltzmann weight $e^{\varphi}$ related to the periodic potential.  Surprisingly, we find that this relation is not unique, and there are many periodic potentials, which can yield the same Boltzmann weight.  With this idea we study the Brownian motion in various potentials,  in which the Boltzmann weight can be determined either analytically or numerically and confirm these results by numerically solving the Langevin equation.  These results can greatly broaden our understanding of Brownian motion with time-dependent potentials.

We illustrate these results by considering  the following Langevin equation \cite{langevin1908theorie} in the overdamped regime
\begin{equation}
\gamma \dot{x} = f(x) + \xi(t),
\end{equation}
where $f(x) = -dU/dx$, with $U(x)$ being a periodic potential and $\xi$ is the random force satisfying  $\langle \xi(t) \rangle =0$ and $\langle \xi(t) \xi(t') \rangle =2\gamma k_B T \delta(t-t')$, in which the symbol $\langle \cdot \rangle$ means ensemble average. In this equation, the potential $U(x)$ is generally assumed to be a periodic function of $x$, that is, $U(x) = U(x+a)$, where $a$ is the spatial period. The PDF can be written as \cite{SangYang2025Brownian,defaveri2023brownian,sivan2018probability,gong2025proof} 
\begin{equation}
p(x,t) = \exp(-\phi(x)) \mathcal{G}(x, t), \quad \phi =\beta U(x),
\label{eq-pxt}
\end{equation}
where $\beta = 1/k_B T$ is the inverse temperature ($\beta=1$ in the following) with $k_B$ denoting the Boltzmann constant, and $\mathcal{G}$ is a Gaussian envelope 
\begin{equation}
\quad \mathcal{G}(x, t) = {1\over \sqrt{4\pi D^*t}} \exp(-{(x-x_c)^2 \over 4D^* t}),
\label{eq-gauss}
\end{equation}
where $D^*$ is the effective diffusion constant and $x_c$ is the center of the envelope. However, this problem has not been studied with time-dependent potential, that is, $U = U(x, t)$. Thus we have to address the fundamental question that to what extent and how the above PDF can be the universal asymptotic solution of the corresponding Fokker-Planck  equation with time-dependent periodic potentials? 

{\it LJ formula and effective diffusion constant $D^*$}.  We first build an intuitive picture to understand the LJ formula in static periodic potentials. For the given  Langevin equation $\gamma \dot{x} = -U_x(x, t) + \xi$, the PDF is determined by the following Fokker-Planck equation \cite{smoluchowski1916über,Fokker1914Die,Planck1917Sitzungsberichte,Risken1996The}
\begin{equation}
{\partial p \over \partial t} = D_0 {\partial \over \partial x} e^{-\phi} {\partial \over \partial x} e^{\phi} p,
\end{equation}
where now $\phi = \beta U(x, t)$ and $p = p(x, t)$. The previous researches have mainly focused on the effective diffusion constant \cite{Gang1996Diffusion,Marchenko2018Enhanced,Marchenko2023Temperature,Marchenko2025Giant} in this model, yielding the LJ  formula for static periodic potentials \cite{lifson1962self,Festa1978Diffusion,gunther1979mobility,weaver1979effective}
\begin{equation}
D^* = \lim_{t\rightarrow \infty} {\sigma(t) \over 2t} = {D_0 \over \langle e^{\phi}\rangle_x \langle e^{-\phi}\rangle_x},
\label{eq-LJ}
\end{equation}
in which the mean Boltzmann weight is defined as $\langle \exp(\pm\phi)\rangle_x = {1\over a} \int_0^a dx \exp(\pm \phi)$, and $\sigma(t) = \langle x^2\rangle - \langle x\rangle^2$ is the variance of position while $D_0$ is the diffusion constant without external potential, thus is determined by $\gamma \dot{x} = \xi$. This is a standard result from the diffusion equation with Gaussian function given by $\mathcal{G}$. For sake of self-consistent, we first briefly summarize the major idea in Ref. \cite{SangYang2025Brownian}.  We can assume $p = \exp(-\phi) \mathcal{G}$, where $\exp(-\phi)$ is the Boltzmann weight,  yielding 
\begin{equation}
    \exp(-\phi) {\partial \mathcal{G} \over \partial t} = D_0  {\partial \over \partial x} e^{-\phi} {\partial \over \partial x} \mathcal{G}.
    \label{eq-Gx_eq}
\end{equation}
Using the periodicity of $\phi$ and assume $\mathcal{G}$ changes very small in a full period $a$, we can take its average in a full period, yielding $\langle \exp(-\phi)\rangle_x  {\partial \mathcal{G} \over \partial t} = D_0  e^{-\phi} {\partial^2 \over \partial x^2} \mathcal{G}$. Then we move $e^{-\phi}$ to the left side and applying the same procedure again, we can obtain an effective diffusion equation that $
{\partial \mathcal{G} \over \partial t} = D^*  {\partial^2 \mathcal{G} \over \partial x^2}$, with $D^*$ given by the LJ  formula.

{\it Partial equation for the Boltzmann weight}: The above procedure, obviously, is not applicable to Brownian motion in time-dependent potentials.  Let us assume $p = e^{\varphi} \mathcal{G}$,  where $e^{\varphi}$ is the Boltzmann weight, as given by Eq. \ref{eq-varphi}. Plug this 
expression into the Fokker-Planck equation and we obtain 
\begin{align}
\varphi_t + \partial_t \ln \mathcal{G} = &D_0[(\phi_{xx} + \varphi_{xx}) + (\phi_{x} + \varphi_{x}) \varphi_x] \notag \\
&+ D_0[(\phi_x+2 \varphi_x)\partial_x\ln \mathcal{G} + \mathcal{G}_{xx}/\mathcal{G}]. 
\end{align}
Our key insight is that when $\mathcal{G}$ is a Gaussian envelope, the direct coupling between the Boltzmann weight and the Gaussian profile should be decoupled in some way when $t$ is large enough due to their totally different amplitudes,  thus we should have 
\begin{equation}
\varphi_t = D_0\left(\phi_{xx} + \varphi_{xx} + (\phi_{x} + \varphi_{x}) \varphi_x \right). 
\end{equation}
This is a nonlinear partial equation, which we see the similar form in the Kardar-Parasi-Zhang (KPZ) equation \cite{kardar1986dynamic} and Jacobi-Hamiltonian equation.  Assume the Cole–Hopf transformation $\varphi = \ln w$, we obtain the following linear equation 
\begin{equation}
{\partial w \over \partial t} = D_0{\partial J \over \partial x}, \quad J = {\partial w \over \partial x} + \phi_x w. 
\label{eq-WJphix}
\end{equation}
This corresponds to define $p = w \mathcal{G}$, hence $w \ge 0$ for all $x$ and $t$ for any $\phi_x$ is assumed.  Eq. \ref{eq-WJphix} is the key equation to the Boltzmann weight in the Fokker-Planck equation, bridging the relation between $\varphi$ and $\phi$, which can be determined numerically. The solution of $\mathcal{G}$ is given by 
\begin{equation}
\partial_t \ln \mathcal{G} = D_0(\phi_x+2 \varphi_x)\partial_x\ln \mathcal{G} + D_0\mathcal{G}_{xx}/\mathcal{G}. 
\end{equation}
Defining $\mathcal{U} = \phi + 2 \varphi$, the above equation can be written in the following compact form
\begin{equation}
{\partial \mathcal{G} \over \partial t} = D_0 e^{-\mathcal{U}}{\partial \over \partial x} e^{\mathcal{U}} {\partial \mathcal{G} \over \partial x}.
\label{eq-Gxt}
\end{equation}
This equation is analogous to Eq. \ref{eq-Gx_eq}, yet its asymptotic solution should be related to the Gaussian envelope characterized by $x_c$ and $D^*$ since the Boltzmann weight has been excluded in the PDF. Thus, we expect $\mathcal{G}$ to take the form given in Eq. \ref{eq-gauss}, with $D^*$ and $x_c$ to be determined numerically. This expectation is supported by the well-defined nature of $D^*$ and further verified by numerical solving of Eq. \ref{eq-Gxt} using finite difference method. Further, this Gaussian profile can be understood using the method in Ref. \cite{SangYang2025Brownian}. This example show that the Gaussian profile is the asymptotic solution of some more general differential equations. 

To understand the physics  in the above two equations,  we first assume $\varphi$ (and $w$) to be time-independent, hence
\begin{equation}
{\partial w \over \partial x} + \phi_x w = f(t). 
\end{equation}
where $f(t)$ can be any function of time. When $f(t) = 0$,  we have ${\partial w \over \partial x} + \phi_x w = 0$, yielding $\phi = -\varphi$, up to a constant,  using the relation $\varphi = \ln w$.  This result is  already well-known in the previous literature \cite{sivan2018probability,defaveri2023brownian,SangYang2025Brownian}. We have shown that the periodicity of the potential is not necessary, thus this weight can be generalized to the other potentials,  including the harmonic potentials. When $f(t) \ne 0$ we have 
\begin{equation}
\phi_x = {f(t) \over w} - {\partial \ln w \over \partial x}.
\end{equation}
Furthermore, this result will lead to  the following theorem:
\begin{theorem}
In Eq. \ref{eq-WJphix}, the potential $\phi$
and the potential $\phi + \int dx f(t)/w(x, t)$ should correspond to the same Boltzmann weight $w(x, t)$ with any function $f(t)$. 
\end{theorem}
Yet these different potentials may have different $D^*$ and $x_c$, hence have different PDFs.  This result is a direct consequence of absence of boundary conditions. Meanwhile, it is quite possible that while these potentials $\phi(x, t)$ are time dependent, the Boltzmann weight can be time-independent, that is $w = w(x)$.  More interesting results will be discussed below, which will be verified through numerical simulations of the Langevin equation.

\begin{figure}
    \centering
    \includegraphics[width=1\linewidth]{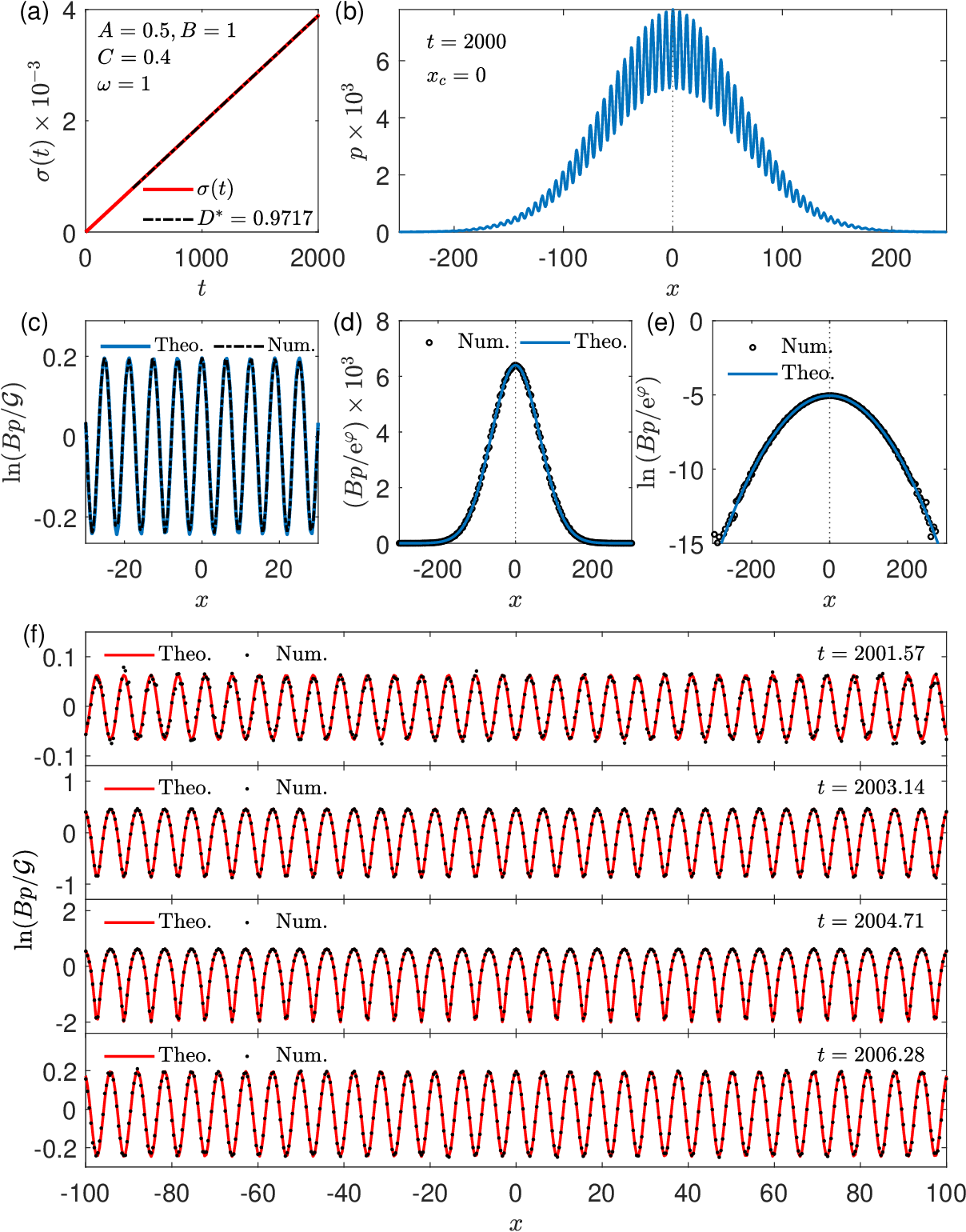}
    \caption{The Brownian motion and PDF  with  $w = B + A \cos(\omega t) \cos(x) +  C \cos(x)$ and $f(t)=0$. (a) The variance $\sigma(t)= \langle x^2\rangle - \langle x\rangle^2$ versus $t$, from which the slope yields the effective diffusion constant $D^*$. (b) PDF at $t = 2000$. (c) $\varphi(x, t) = \ln Bp/\mathcal{G}$ versus $x$ at $t = 2000$ from theory and numerical simulation, without fitting parameters involved. (d) $Bp/e^{\varphi}$, yielding the Gaussian profile. (e) Demonstration of the Gaussian profile in the large $x$ limit.  (f) $\varphi = \ln Bp/\mathcal{G}$ for various $t$.   }
    \label{fig-fig1}
\end{figure}

{\it An exact solvable solution}:  We first illustrate the above surprising results using an exact solvable model.  For a given $\phi$, it is generally very hard to obtain the solution of $w$ analytically. Our strategy is to obtain $\phi$ for a given $w$, which should be much more simplified. To this strategy, we use the following function to demonstrate our idea
\begin{equation}
w = B + A \cos(\omega t) \cos(x) +  C \cos(x),
\label{eq-model1_w}
\end{equation}
where we assume $B > |A|+ |C|$ to ensure that $w > 0$ for any $t$ and $x$. In this way, we should have 
\begin{equation}
{\partial w \over \partial x} + \phi_x w = f(t) - \frac{A \omega}{D_0} \sin(\omega t) \sin(x),
\end{equation}
for any time-dependent function $f(t)$. With this expression, we can obtain $\phi_x$ as 
\begin{equation}
\phi_x = \frac{f(t)+ \left[C+A\cos(\omega t)-\frac{A\omega}{D_0} \sin(\omega t)\right]\sin(x)}{B+A \cos(\omega t)\cos(x) + C\cos(x)}, 
\end{equation}
which give the force acting on the particle and we have
\begin{align}
\phi = &\frac{2 \tanh^{-1}\left( \tan(\frac{x}{2} )\sqrt{\frac{ C + A \cos(\omega t) - B }{ C + A \cos(\omega t) +B } } \right) f(t)}{ \sqrt{  (C + A \cos(\omega t) - B) (C + A \cos(\omega t) +B) } } \notag \\
    &+ \frac{A \omega \sin(\omega t) \ln(w) }{ D_0\left[C + A \cos(\omega t)\right] }  - \ln(w) + \phi_0. 
\end{align}
Here the constant $\phi_0$ is not important. Thus we see that for a given $f(t)$, $\phi$ should be totally different; however, all these functions give the same $w$, thus give the same $\varphi$. This is different from the time-independent potential, in which $\varphi = - \phi$ exactly. For this reason, we see that there are many $\phi$,  which yield the same Boltzmann weight; see the above theorem.  This is a new feature in the Fokker-Planck equation not uncovered before. This strategy has an obvious advantage that for these $\phi$, the only parameter need to be determined in our numerical simulation is the effective Diffusion constant $D^*$, which can be fitted through the numerical simulation of the Langevin equation.

We emphasize that in the above model the parameters $A$, $B$ and $C$ can be time independent asymptotically. The reason is that when $t$ is large enough, the Gaussian envelope changes negligibly in a full period, thus we should have $\int_{x}^{x+a} \cos(x) \mathcal{G}(x, t) dx = 0$,  when the width $\sqrt{\sigma(t)}$ is much larger than the period $a$.  Hence for any $A$ and $C$, the PDF  is already normalized. Naturally, $1/B$ is the normalization factor in this model, which should be time-independent. 

Next, we verify these results using numerical method. We solve the  Langevin equation using Euler-Maruyama method and average the results by $5\times 10^7$ to $3 \times 10^8$ times to obtain the PDF $p(x, t)$. The results are presented in Fig. \ref{fig-fig1}. In Fig. \ref{fig-fig1}(a) the variance of position $\sigma(t)= \langle x^2 \rangle - \langle x\rangle^2$ increase linearly with the time $t$,  with $x_c = \langle x\rangle =0$,  yielding a
well defined  diffusion constant  $D^* = \lim_{t\rightarrow \infty} \sigma(t)/2t$ \cite{Marchenko2018Enhanced,Marchenko2023Temperature,Marchenko2025Giant,Gamba2024Hyperballistic}. The PDF is presented in Fig. \ref{fig-fig1}(b), and using $D^*$ and $x_c$, we can determine the Boltzmann weight, and compare its value with the analytical results in Fig. \ref{fig-fig1}(c). The Gaussian envelope, excluding the Boltzmann weight from $p(x, t)$, is presented in details in Fig. \ref{fig-fig1}(d) and (e).  These results except only $D^*$ and $x_c$ demonstrate the validity of our solution.

\begin{figure}
    \centering
    \includegraphics[width=0.9\linewidth]{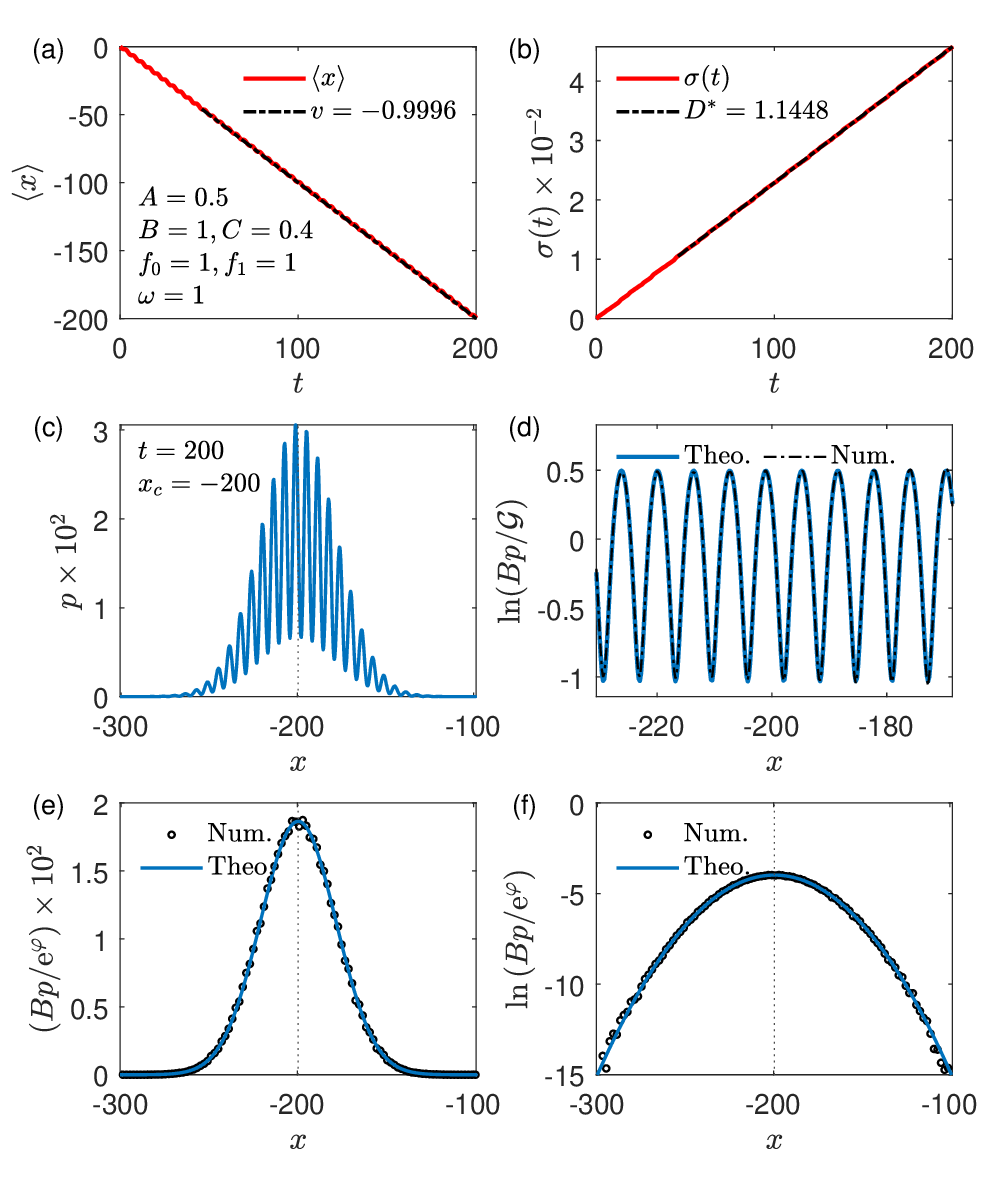}
    \caption{The Brownian motion and PDF with $w = B + A \cos(\omega t) \cos(x) +  C \cos(x)$ and $f(t) = f_0+f_1 \cos(\sqrt{2}t)$. (a) The mean position $x_c = \langle x \rangle$ versus $t$, where the Gaussian envelope is given by Eq. \ref{eq-gauss}. (b) The variance  $\sigma(t)= \langle x^2 \rangle-\langle x \rangle^2$, from which the slope yields the effective diffusion constant $D^*$. (c) PDF at $t = 200$. (d) $\varphi(x, t) = \ln Bp/\mathcal{G}$ versus $x$ at $t = 200$ from theory and numerical simulation, without fitting parameters involved. (e) $Bp/e^{\varphi}$, yielding the Gaussian profile. (f) Demonstration of the Gaussian envelope in the large $x$ limit.}
    \label{fig-fig2}
\end{figure}

Further, let's consider a more interesting case when $f(t) = f_0 + f_1 \cos(\sqrt{2}t)$, in which the Boltzmann wight $w(x,t)$ will have the same form as the case in Fig. \ref{fig-fig1}. For this model, we find that the center of the Gaussian envelope propagates with a velocity $v=  -f_0/B$, which is obtained by our numerical simulation. The results are presented in Fig. \ref{fig-fig2}(a) and (b). Similarly, we present the PDF at $t = 200$ in Fig. \ref{fig-fig2}(c) and compare the numerical result of Boltzmann wight with the theoretical prediction in Fig. \ref{fig-fig2}(d), which agree with high accuracy. In Fig. \ref{fig-fig2}(e) and (f) we demonstrate the nature of Gaussian profile.

\begin{figure}
    \centering
    \includegraphics[width=0.9\linewidth]{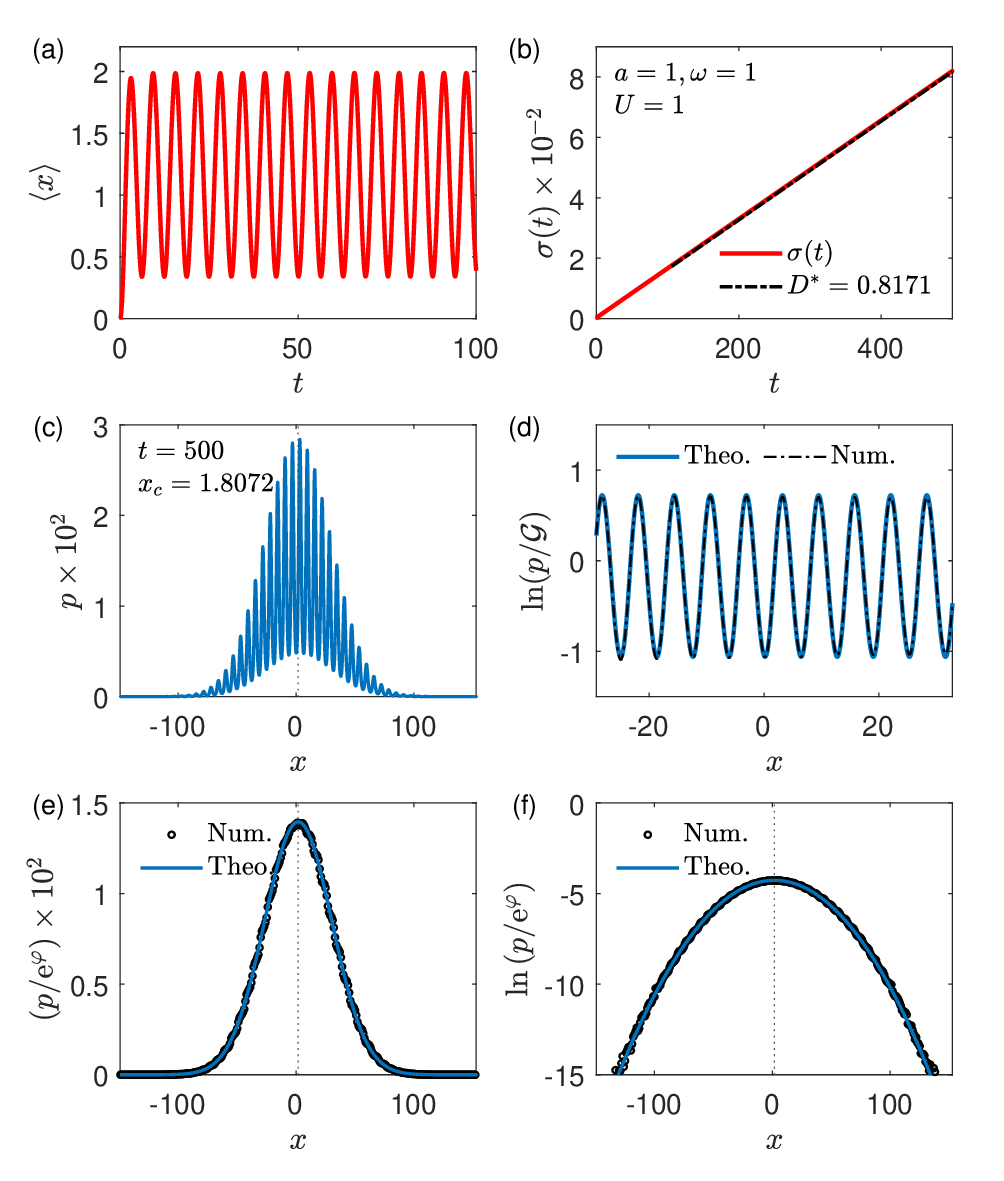}
    \caption{The same as Fig. \ref{fig-fig2} with  $\phi = U \cos(x) - a \sin(\omega t) x$. The PDF in (c)-(f) consider a Gaussian envelope $\mathcal{G}(x-x_c,t)$, where $x_c = \langle x \rangle$ at $t = 500$ as shown in (a). The theoretical $\varphi(x,t)$ in (d) is computed by $\varphi = \ln w$, where $w$ is the solution to Eq. \ref{eq-WJphix} obtained using finite difference method.}
    \label{fig-fig3}
\end{figure}

{\it Boltzmann weight in general potentials from Eq. \ref{eq-WJphix}}: For the general potentials $\phi(x, t)$, we can not obtain the Boltzmann weight exactly. Instead, we can solve the Fokker-Plank equation using two different steps. Firstly, we can determine the effective diffusion constant $D^*$ and the center of the Gaussian envelope $x_c(t)$, based on which we can determine the Gaussian envelope. Next, we solve Eq. \ref{eq-WJphix} exactly to determine the Boltzmann weight $w(x, t)$. To be specific, we consider the following time dependent periodic potential \cite{Marchenko2018Enhanced,Marchenko2023Temperature,Marchenko2025Giant} 
\begin{equation}
\phi = U \cos(x) - a \sin(\omega t) x,
\end{equation}
in which $F = -\phi_x = U \sin(x) + a \sin(\omega t)$ represents the time-dependent force acting on the particle. It has been found for a long time that $D^*$ is well defined in various conditions, indicating that the overall profile should be related to the Gaussian envelope in some way.We solve the partial differential equation Eq. \ref{eq-WJphix} using finite difference method, and find that when $t$ is large enough, the initial function $w(x, 0)$ is not important anymore. Our results are presented in Fig. \ref{fig-fig3}, in which Fig. \ref{fig-fig3}(a) and (b) gives the mean position $x_c$ and the diffusion constant. In Fig. \ref{fig-fig3}(c) we present the PDF obtained by solving the Langevin equation. In Fig. \ref{fig-fig3}(d), we present the weight function determined theoretically by solving Eq. \ref{eq-WJphix}, showing of excellent agreement with that obtained from the Langevin simulation using $p(x, t)$ and $\mathcal{G}(x, t)$. The Gaussian envelope is verified in Fig. \ref{fig-fig3}(e) and (f). 

\begin{figure}
    \centering
    \includegraphics[width=1\linewidth]{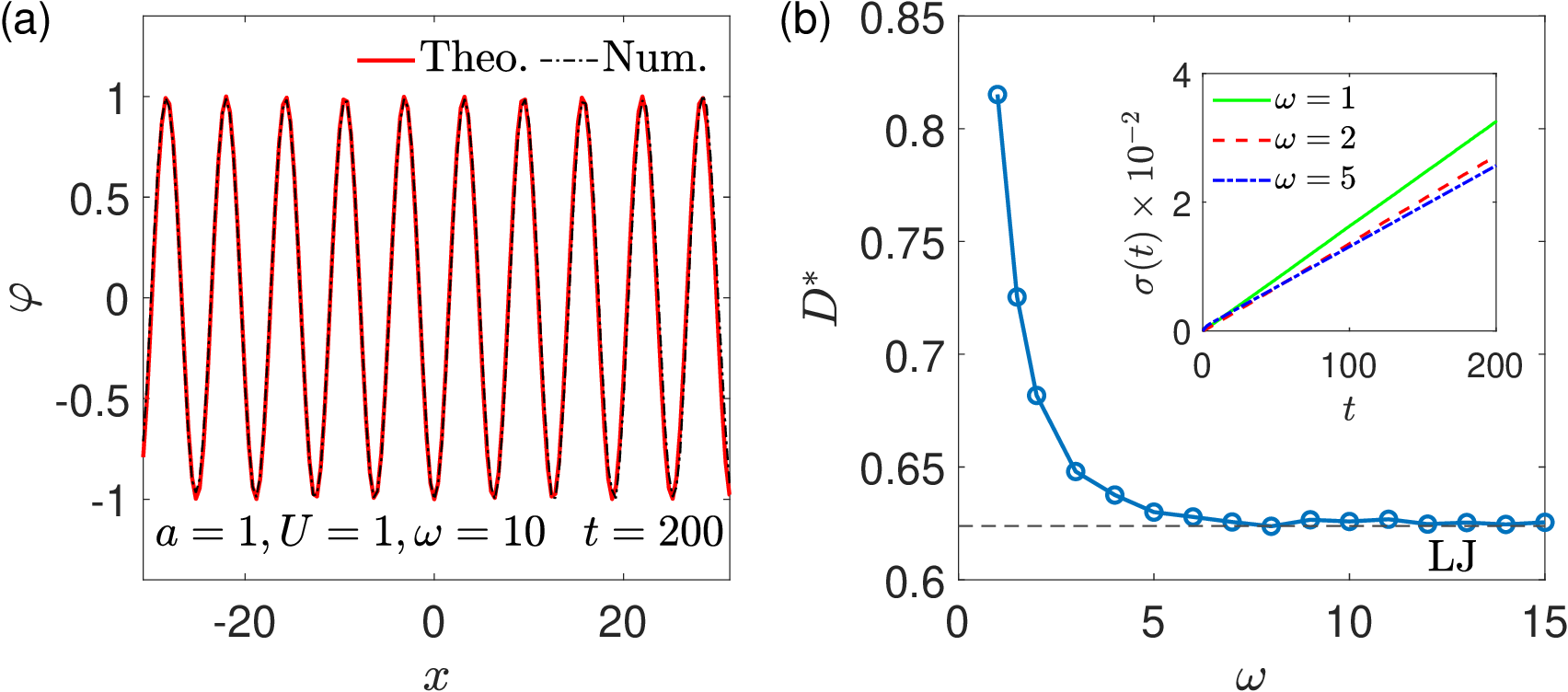}
    \caption{The potential $\varphi$ and effective diffusion constant $D$ in the high frequency limit. (a) The theoretical and numerical result of $\varphi(x,t)$ at $t= 200$, with $\omega =10$. (b) $D^*$ and variance of position (inset) as a function of $\omega$. The black dashed line represents $D^*$ calculated using LJ formula. }
    \label{fig-fig4}
\end{figure}

Accordingly, this model will have an interesting limit that when $\omega$ is large enough, the Boltzmann weight should approach a stable solution that $w(x, t)$ will become time-independent. In this way, we will find that $\varphi = -U\cos(x)$, ignoring the fast modulating term in the phase. Meanwhile $D^*$ will also quickly approach the LJ limit calculated by Eq. \ref{eq-LJ}. These results are presented in Fig. \ref{fig-fig4}. Obviously, when $U = 0$ and $\omega$ is large enough, the Langevin equation $\gamma \dot{x} = a \cos(\omega t) x + \xi$ can be solved exactly, yielding the limit $\gamma \dot{x} = \xi$, with $D^* = D_0$. 

{\it Entropy extremum}: With this PDF we can determine the entropy \cite{Aghion2019From, Aghion2020Infinite}
\begin{equation}
S(t) = -\langle \ln p\rangle = {\sigma(t) \over 4D^*t} +{1 \over 2} \ln (4\pi D^* t) + {E_\text{p} \over T},
\end{equation}
where $k_B =1$ and  $E_\text{p} =- \langle \varphi\rangle/\beta$ is the effective mean potential energy.  In this way, we expect the thermodynamic law that $\partial S(t)/\partial E_\text{p}  = 1/T$, even though $E_\text{p}$ is time-dependent. Furthermore, it is quite possible that $S(t)$ is not a monotonic increasing function of $t$, yet in the long time limit, the entropy increases logarithmically with time given by $\ln (4\pi D^* t) /2$. Following \cite{Aghion2019From, Aghion2020Infinite},  this PDF can be understood in terms of entropy extremum principle \cite{Janyes1947Information} using  $\varphi$, instead of $\phi$. 

{\it Conclusion and Discussion}: To conclude, we present a much straightforward way to obtain the universal asymptotic solution of the Fokker-Planck equation in the over damped Langevin equation. We show that the PDF can be divided into a Gaussian envelope $\mathcal{G}$ characterized by the effective diffusion constant $D^*$ and the center $x_c$ multiplied by a Boltzmann weight $w$. We derive the equation for the Boltzmann weight  and find that there are many different potentials, which can yield the same weight function. These results are confirmed by solving the Langevin equation numerically. 
We show that the entropy of this PDF satisfies the thermodynamic law. 

This result has some immediately applications. Firstly, it can be generalized to higher dimensions \cite{Speer2009Directing}. For the two-dimensional periodic potentials when $D^*$ is well defined in terms of Einstein relation, the Boltzmann weight related to the potential should be given by ${\partial w \over \partial t} = \nabla \cdot {\bf J}$, where ${\bf J} = \nabla w + (\nabla \phi) w$, with $\nabla = (\partial_x, \partial_y)$. This is  a direct generalization of Eq. \ref{eq-WJphix}. Furthermore,  these results are significant in designing and understanding of Brownian motion in  different platforms utilizing Josephson junctions \cite{Ambegaokar1969Voltage,Devoret1987Resonant,Berthold2017Quantum,Trahms2023Diode,Daviet2023Nature}, rotor models \cite{Bellando2022Giant,Chaoxiong2025Kinetic} and even artificial periodic structures \cite{Martin1997Brownian,hanggi2009artificial},  ultracold atoms \cite{Kindermann2017Nonergodic,Wang2025Observation} and active matters \cite{Elgeti2015Physics,Bechinger2016Active,Basu2018Active,Löwen2020Inertial,Antonov2024Inertial,Maggi2014Generalized}. In the presence of  inertia term \cite{Marchenko2018Enhanced,Marchenko2023Temperature,Marchenko2025Giant,Goychuk2021Nonequilibrium}, the similar PDF should be observed, yet the above equations should be modified in accordingly.  Finally, our results are expected to be generalized to time-dependent quasi-periodic boundaries \cite{gong2025proof,SangYang2025Brownian} and trapped potentials (set $\mathcal{G} =1$) 
\cite{Tashiro2007Brownian,Chakraborty2012Persistence,Ryabov2015Brownian}.  Thus the asymptotic  solution of the Fokker-Planck equation presented in this work should open a new field for the searching of Brownian motion in more  physical systems \cite{Giordano2024Effective,Guo2023Brownian,Leptos2009Dynamics,Bia2020Colossal} and explore their thermodynamics \cite{Aghion2019From, Aghion2020Infinite}, which should also be some kind of interests in probability theory in applied mathematics \cite{Chhib2004Influence,Ji2021Convergence,Alpatov1994Spectral}. 

%. We may solve this equation using numerical yield, which will yield much rich physics. This is similar to the Liuvioieu equation, yet it is not, because $w$ is not the probaility distribution. 
% 要强调不是刘维尔方程。
% 讨论可能的应用
% 2d
% 非周期系统，准周期系统
% 具有惯性的系统
% 反常扩散系统，将$t$替换为$t^* = t^\alpha$，这样可以得到一些奇怪的结论。
% 实现的问题，可以哪些实验来实现。

% 是否要讨论/强调，多对1，是这个方程的最基本的特点，隐藏很深的基本特点。这是我们揭露出来的一个重要特点。所以值地仔细研究。这个东西，和gauge potential是否有关？好像无关。但是又似乎有一些联系。有很多不同的phi，对同一个varphi。

\bibliography{ref.bib}

\end{document}